\pdfoutput=1
\documentclass[slac_one]{revtex4}
\usepackage{txfonts}
\usepackage{graphicx}
\usepackage{fancyhdr}
\newcommand{\D}{\ensuremath{D}}
\newcommand{\Dbar}{\ensuremath{\overline{D}}}
\newcommand{\Ds}{\ensuremath{D_{s}}}
\newcommand{\Dsp}{\ensuremath{D_{s}^+}}
\newcommand{\Dsm}{\ensuremath{D_{s}^-}}

\newcommand{\Dz}{\ensuremath{D^0}}
\newcommand{\Dzb}{\ensuremath{\overline{D}^0}}
\newcommand{\Dp}{\ensuremath{D^+}}
\newcommand{\Dm}{\ensuremath{D^-}}
\newcommand{\Kp}{\ensuremath{K^+}}
\newcommand{\Km}{\ensuremath{K^-}}
\newcommand{\KS}{\ensuremath{K^0_{S}}}

\newcommand{\pip}{\ensuremath{\pi^+}}
\newcommand{\pim}{\ensuremath{\pi^-}}
\newcommand{\piz}{\ensuremath{{\pi^0}}}

\newcommand{\invpb}{pb$^{-1}$}

\newcommand{\cleoc}{\mbox{CLEO-c}}
\newcommand{\Br}{\ensuremath{\mathcal{B}}}

\preprint{EFI 08-28}
\begin{document}

\title{Hadronic Charm Decays: An Experimental Review} 

%

\author{P.~U.~E.~Onyisi}
\affiliation{Enrico Fermi Institute, University of Chicago, Chicago, IL 60637, USA}

\begin{abstract}
I review some recent results on hadronic decays of charmed mesons.  These decays illuminate a wide range of physics, including the absolute normalizations of $b$ and $c$ decays, the understanding of \Dz\ mixing, the coupling of heavy mesons to hadronic decay products, and light hadron spectroscopy. 
\end{abstract}

\maketitle

\thispagestyle{fancy}


\section{INTRODUCTION} 
The lightest open charm mesons --- the \Dz, \Dp, and \Dsp\ --- decay through the weak interaction, and the majority of their decays are to final states containing only hadrons.  These mesons and their hadronic decays lie at an intersection of weak and strong physics: the easiest detection of charmed states for flavor physics is through hadronic decays with large rates and simple topologies, while the decay processes themselves provide important information on hadronic spectroscopy and strong interactions.

Four groups of measurements are discussed here.  Two of them, absolute branching fractions of $D$ mesons and strong phases in hadronic decays, are important for allowing precision studies of the CKM matrix and other physics to reduce systematic uncertainties.  The other two, the studies of two-\ and three-body $D$ decay amplitudes, shed light on how long distance hadronic physics affects the visible results of short-distance weak processes.

\section{EXPERIMENTS}
Recent activity in the field has been dominated by the charm experiment \cleoc\ and the $B$-factories BaBar and Belle.  Detailed understanding of hadronic charm decays places a premium on efficient and pure particle identification, high resolution on daughter hadron momenta, well-understood initial states, and large datasets, which these experiments can provide. 

The \cleoc\ experiment at the CESR symmetric $e^+ e^-$ storage ring took data in the center of mass energy range between 3.67 and 4.26 GeV to investigate topics in both open charm and charmonium physics.  Primary data collection in the charm region occurred between 2003 and 2008.  Of particular relevance here are 818 \invpb\ of $\psi(3770)$ data for \Dz\ and \Dp\ studies at threshold and around 600 \invpb\ in the region of 4.17 GeV collected for \Dsp\ physics.  This yielded approximately 3 million $\Dz\Dzb$, 2.4 million \Dp\Dm, and six hundred thousand $\Ds^{*\pm}\Ds^\mp$ pairs.  By running just above threshold for the production of various species, it could be guaranteed that \Dz, \Dp, or \Dsp\ mesons would always be produced with their antiparticles.  This is one of the most prominent features of the \cleoc\ dataset and enables powerful partial reconstruction (or ``tagging'') techniques, as the typical efficiency for full reconstruction of a single \D\ meson (in one of a menu of clean decay modes) is between 5 and 10\% and the presence of the antiparticle is implied for those events.  In addition, the low final state multiplicity of $D$ and continuum light quark production processes at this energy means combinatoric background is generally small.

The BaBar and Belle experiments are detectors at the PEP-II and KEKB asymmetric $e^+ e^-$ facilities, respectively.  Both experiments began taking data in 1999; Belle is ongoing, while BaBar's run concluded in 2008.  Although the datasets of the experiments are collected in the $B\overline{B}$ threshold region, there is an appreciable rate for charm production, both from $B$ decays and from continuum $e^+ e^- \to c\bar c$ events.  The analyses mentioned here use datasets ranging from 210 to 550 fb$^{-1}$.  The large luminosities compensate for the relatively lower efficiencies compared to \cleoc, and statistical precision can be comparable or better at the $B$-factories.  The boost of the center of mass frame at these experiments also allows them to reduce backgrounds by taking advantage of displaced vertices due to $D$ meson lifetimes.

\section{ABSOLUTE BRANCHING FRACTIONS}
Certain high rate, low multiplicity decays of $D$ mesons serve as standard calibrations for decays of charm and bottom quarks.  With very few exceptions, all measurements of branching fractions or decay widths for $D$ or $B$ mesons depend eventually on these reference decays.  For example, measurements of $|V_{cb}|$ via exclusive $B \to D^{(*)} \ell \nu$ decays require knowing the absolute width for this process, which requires precision knowledge of \Dz\ and \Dp\ decay rates.

\begin{figure}
 \includegraphics[width=.48\linewidth]{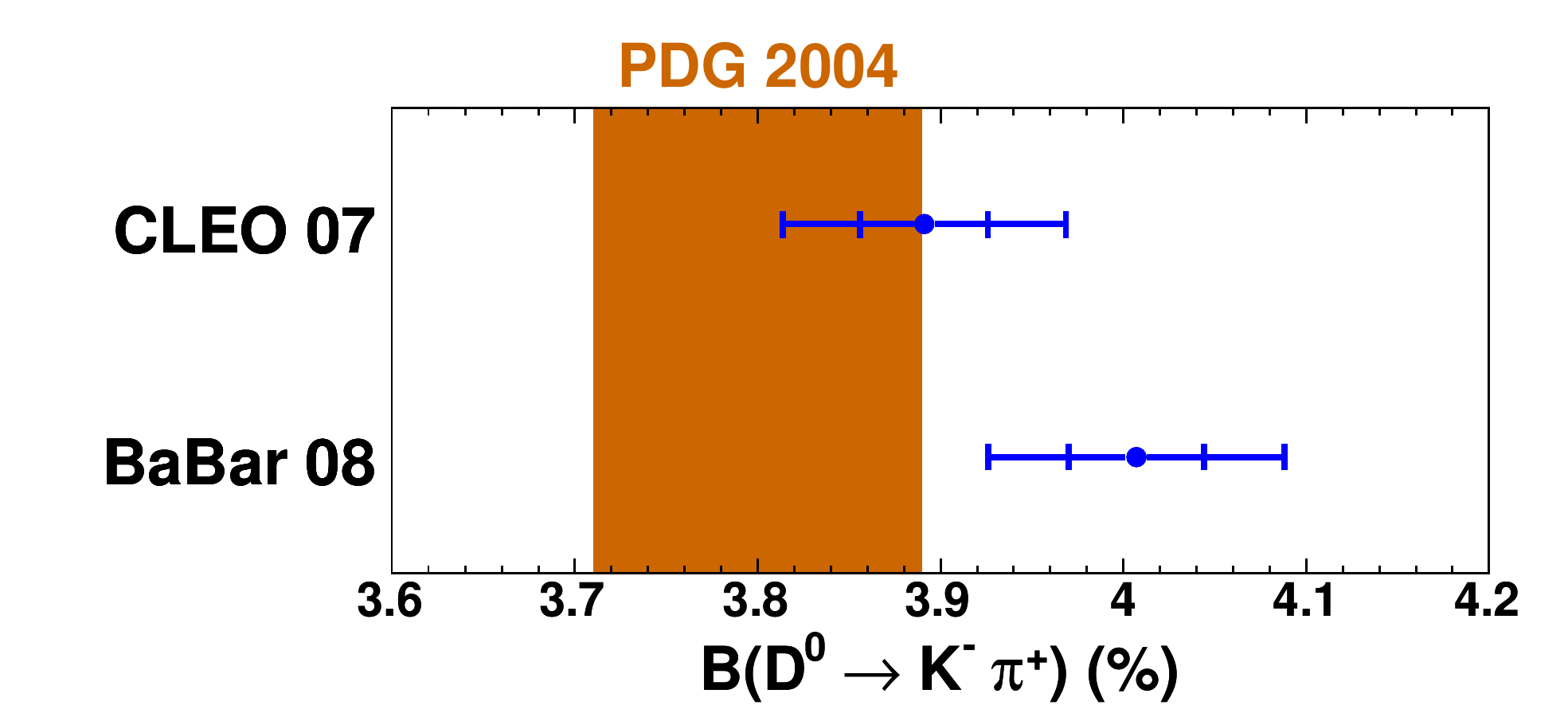}
 \includegraphics[width=.48\linewidth]{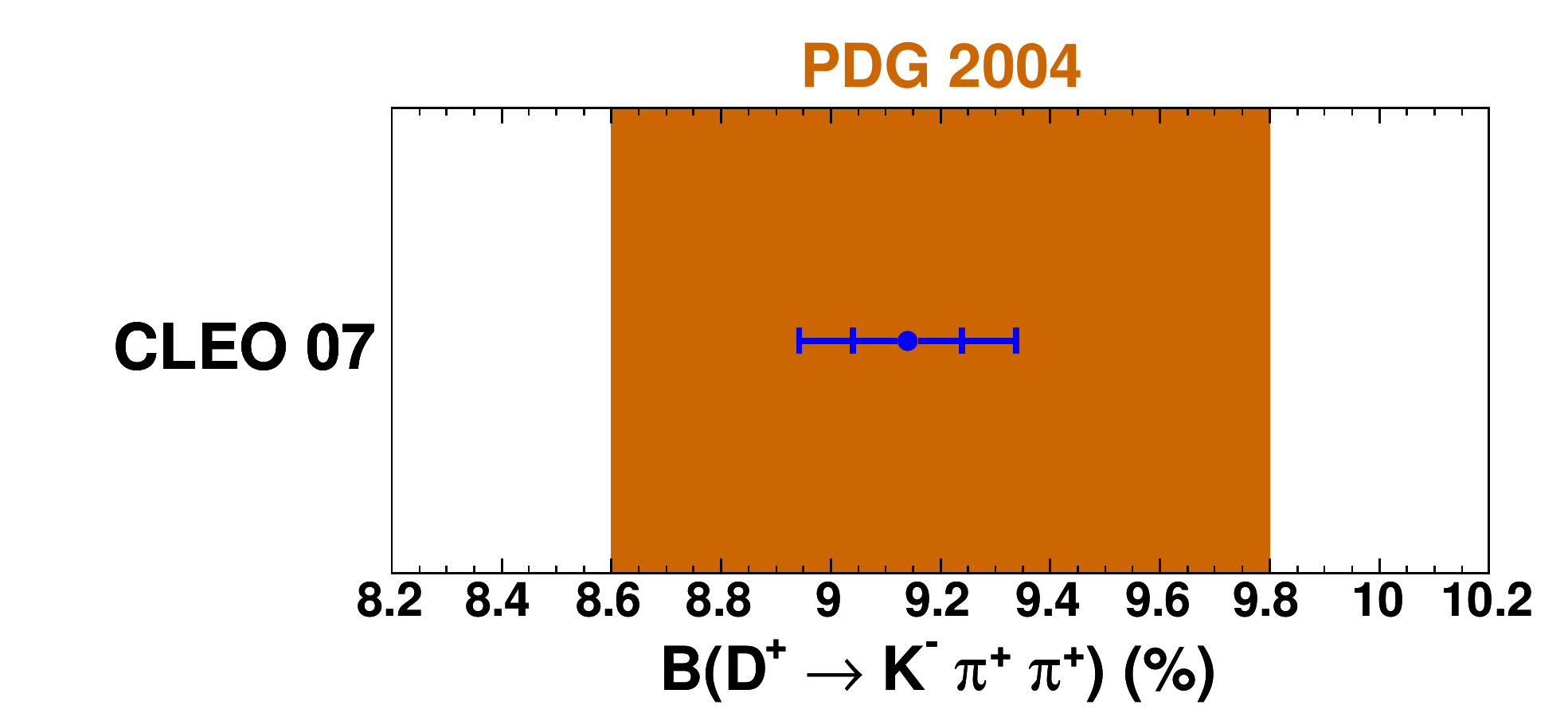}\\
 \includegraphics[width=.48\linewidth]{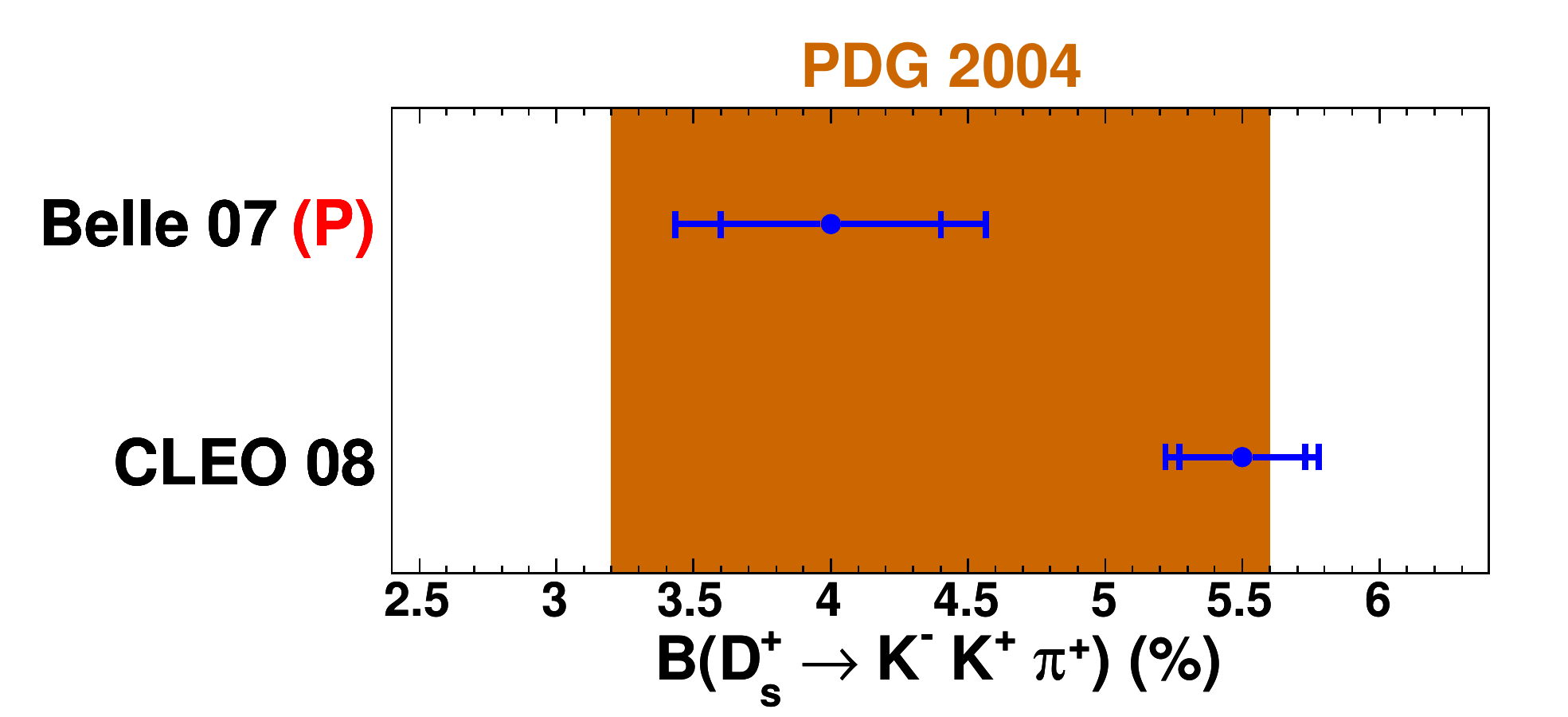}
\caption{\label{fig:dbfs}Measurements of the branching fractions of three reference $D$ decay modes: (top left) $\Dz\to\Km\pip$, (top right) $\Dp\to\Km\pip\pip$, (bottom) $\Dsp\to\Km\Kp\pip$.   Ref.~\cite{Dobbs:2007zt} is ``CLEO 07''; Ref.~\cite{Alexander:2008cqa} is ``CLEO 08''; Ref.~\cite{Aubert:2007wn} is ``BaBar 08''; and Ref.~\cite{Abe:2007jz} is ``Belle 07.''  The comparison is made to the PDG 2004 average \cite{Eidelman:2004wy} as that is the last edition not to include new input from CLEO, BaBar, or Belle.}
\end{figure}

Over the last few years there has been significant improvement in our knowledge of these rates, as illustrated in Fig.~\ref{fig:dbfs} and Table~\ref{tbl:dbfs}.  Between the 2004 and 2008 editions of the Review of Particle Physics \cite{Eidelman:2004wy,Amsler:2008zz}, the uncertainties on the three reference branching fractions shown in Table~\ref{tbl:dbfs} have been improved by factors of 2 to 4.

\begin{table}
\caption{\label{tbl:dbfs} Branching fractions for three important reference decays of \D\ mesons, showing the evolution of between 2004 and 2008.  All branching fractions are in percent.}
 \begin{tabular}{lccc}
\hline\hline
  &$\Br(\Dz\to\Km\pip)$ & $\Br(\Dp\to\Km\pip\pip)$ & $\Br(\Dsp\to\Km\Kp\pip)$\\
\hline
PDG 2004 fit \cite{Eidelman:2004wy} & $3.80 \pm 0.09$ & $9.2 \pm 0.6$ & $4.4\pm 1.2$\\
\hline
\cleoc\ & $3.891 \pm 0.035 \pm 0.069$ \cite{Dobbs:2007zt} &
$9.14\pm 0.10 \pm 0.17$ \cite{Dobbs:2007zt} &
$5.50 \pm 0.23 \pm 0.16$ \cite{Alexander:2008cqa} \\
BaBar & $4.007 \pm 0.037 \pm 0.072$ \cite{Aubert:2007wn} & --- & ---\\
Belle & --- & --- & $4.0 \pm 0.4 \pm 0.4$ \cite{Abe:2007jz} (prel.)\\
\hline
PDG 2008 fit \cite{Amsler:2008zz} & $3.89 \pm 0.05$ & $9.22 \pm 0.21$ & $5.50 \pm 0.28$\\
\hline\hline
 \end{tabular}

\end{table}

\cleoc\ obtains yields of ``single tag'' (a \D\ or \Dbar\ decay is reconstructed in a given mode) and ``double tag'' (both a \D\ and a \Dbar\ are found in specified modes) events, and compares single and double tag yields for various decays to extract the branching fractions.  For $N$ decay modes, yields are obtained for $2N$ single tag decays (\D\ and \Dbar\ are considered separately) and for $N^2$ double tag decays.  The yields can be predicted in terms of $N$ branching fractions, the total number of $\D\Dbar$ pairs, and the detection efficiencies, the last being determined from Monte Carlo simulations.  There are $N^2 + 2N$ measurements and $N+1$ parameters, providing an overconstrained and well-conditioned set of equations which are solved for the branching fractions using either $\chi^2$ minimization or likelihood maximization.  This general technique has been used for \Dz\ and \Dp\ at 3.77 GeV$/c^2$ \cite{Dobbs:2007zt} and for \Dsp\ at 4.17 GeV$/c^2$ \cite{Alexander:2008cqa} with three, five, and eight decay modes respectively.

BaBar measures $\Br(\Dz\to\Km\pip)$ using a partial reconstruction technique \cite{Aubert:2007wn} with the high-rate process $B^0 \to D^{*+} (X) \ell^- \nu$ ($X$ represents possible extra particles that are not reconstructed).  The soft pion from the $D^{*+}$ decay is reconstructed, as well as the lepton $\ell^-$.  The $D^{*+}$ momentum is estimated using the fact that the energy release in $D^{*+} \to \Dz\pip$ is very small so there is a strong correlation between the momenta of the pion and the $D^{*+}$.  The inclusive yield of events is determined from a fit to the implied neutrino mass, assuming the initial $B$ was at rest.  Signal $\Dz\to\Km\pip$ candidates are then reconstructed in these events.  Backgrounds are obtained using below-resonance data for continuum contributions and Monte Carlo simulation of $B\overline{B}$ decays; the background expectations are verified using events where the soft pion and the lepton candidates have the same sign instead of opposite charges.  The result is $\Br(\Dz \to\Km\pip) = (4.007 \pm 0.037 \pm 0.072)\%$.

Belle has a preliminary result for $\Br(\Dsp\to \Km\Kp\pip)$ that uses a double-partial-reconstruction technique \cite{Abe:2007jz}.  They exploit continuum production of $D_s^{*+} D_{s1}^{-}(2536)$ (along with the charge conjugate), with $D_s^{*+} \to \Dsp \gamma$ and $D_{s1}^- \to \overline{D}^* K$.  Four-momentum conservation can be used to infer the presence of the $\Ds^+$ or $\overline{D}^*$ if the other three particles are reconstructed.  The observed yield of $\overline{D}^*K\gamma(\Dsp)_\mathrm{inferred}$ and $\gamma \Dsp K (\overline{D}^*)_\mathrm{inferred}$ depend on the branching fractions of $\Dbar^*$ and $\Dsp$, respectively, and since the total yield of $D_s^{*+} D_{s1}^{-}(2536)$ is the same in both cases the ratio gives information on \Dsp\ branching fractions.  The result is $\Br(\Dsp\to \Km\Kp\pip) = (4.0 \pm 0.4 \pm 0.4)\%$.

As is evident from Table~\ref{tbl:dbfs}, at the moment measurements of \Dz\ and \Dp\ branching fractions are systematics-limited (generally by determination of tracking and particle identification efficiencies) while \Dsp\ measurements are still statistically limited.  To round off the set of important charm hadron reference branching fractions, it would be very desirable to see a result using similar techniques for $\Br(\Lambda_c^+ \to p \Km \pip)$, for which no model-independent measurements currently exist.

\subsection{Normalization of \Dsp\ Decays}
The older BaBar results on $\Br(\Dsp\to\phi\pip)$ \cite{Aubert:2005xu,Aubert:2006nm} have so far not been mentioned.  These measurements impose specific mass requirements on the kaon pair in the $\Km\Kp\pip$ final state.  Because there is a non-negligible contribution of other processes to the $\Km\Kp\pip$ signal in this mass region (notably $\Dsp\to f_0(980)\pip$) the choice of mass window will affect the obtained result.  These results are hard to relate to those using the full $\Km\Kp\pip$ phase space because, in particular, the ratio of branching fractions will {\em not} be the fit fraction of $\Dsp\to\phi\pip$ obtained from an amplitude analysis.  It is worth noting that the 2008 edition of the PDG no longer couples $\Br(\Dsp\to\phi\pip)$ to $\Br(\Dsp\to\Km\Kp\pip)$ via a fit fraction for this reason.

For many experimental purposes, however, it is desirable to restrict the allowed phase space to the $\phi$ region to improve the ratio of signal to background.  There are two proposals to relate branching fractions in restricted and full phase space regions.

One approach was taken in Ref.~\cite{Dobbs:2007zt}, which gave, in addition to the full branching fraction, ``partial'' branching fractions which are obtained by requiring events to have $\Km\Kp$ mass within fixed windows around $m_\phi$.  Except as necessary to derive reconstruction efficiencies, no assumptions are made about the resonant structure in this region --- in particular the measured branching fraction is not corrected for $\Br(\phi \to \Km\Kp)$ and the scalar contribution is fully included.

\begin{figure}
 \includegraphics[width=.33\linewidth]{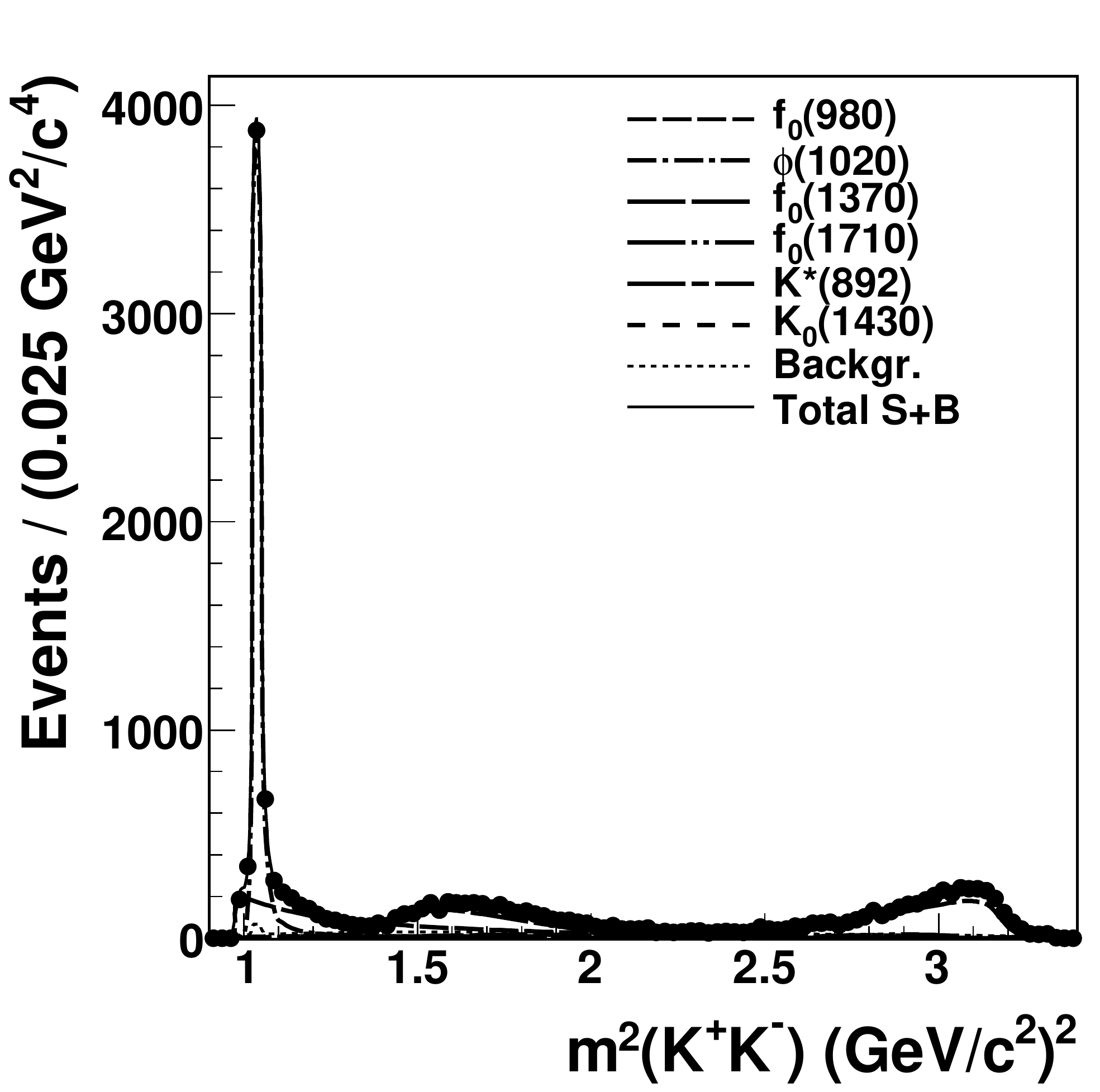}
 \includegraphics[width=.33\linewidth]{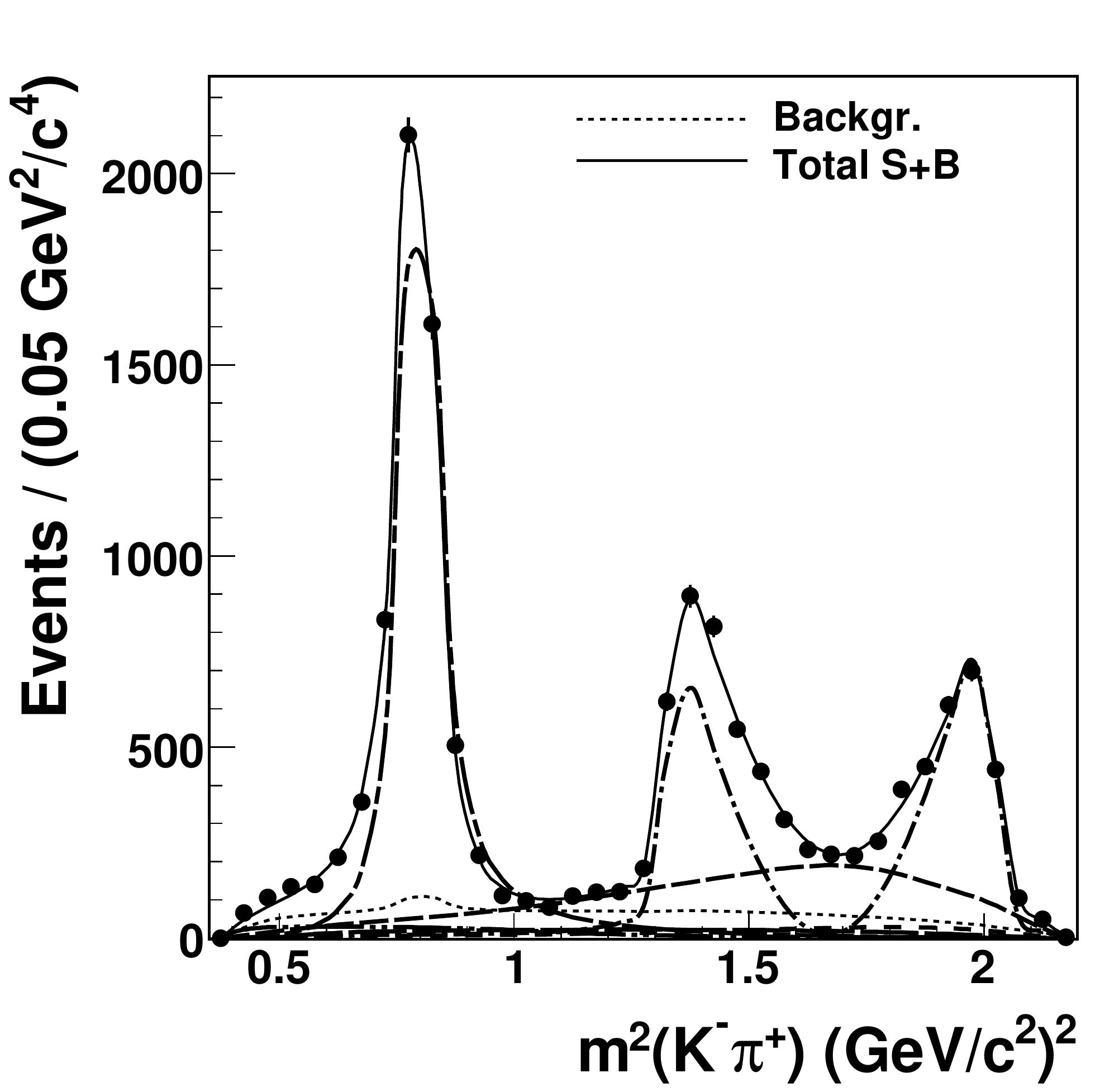} 
 \includegraphics[width=.33\linewidth]{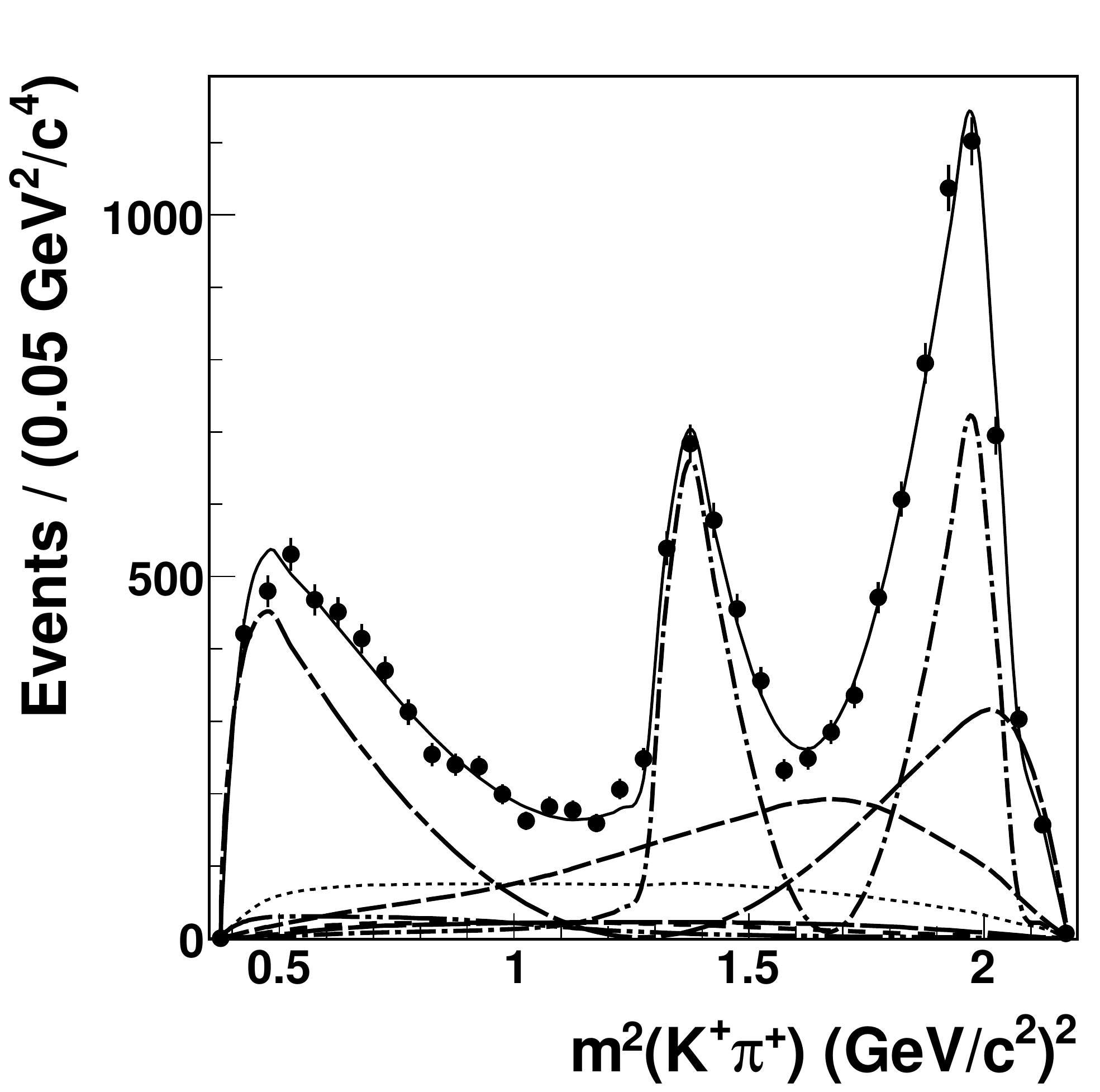}\\
\includegraphics[width=\linewidth]{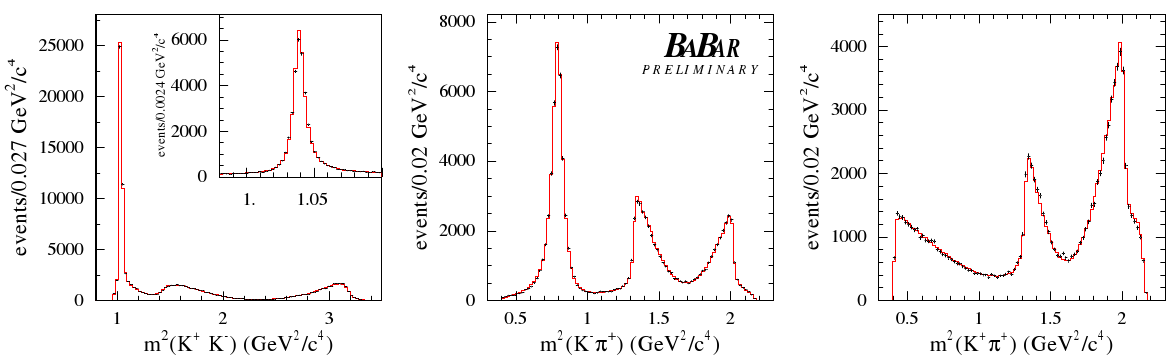}
\caption{\label{fig:kkpidalitz}Projections of data and amplitude fits onto Dalitz varables for the process $\Dsp \to \Km\Kp\pip$ from preliminary CLEO (top) and BaBar (bottom) \cite{Pappagallo:2007it} analyses.  The CLEO analysis uses $14 \times 10^3$ events, and the BaBar result uses $67\times 10^3$ decays.}
\end{figure}

The other, longer term approach is to perform careful amplitude analyses to produce a precise description of the shape of the $\Dsp\to\Km\Kp\pip$ distribution over the entire Dalitz plot.  Other experiments can then implement this description in physics event generators such as EvtGen \cite{Lange:2001uf}, and determine the efficiency of their specific selection choices relative to the full phase space by Monte Carlo simulation.  Towards this end BaBar \cite{Pappagallo:2007it} and CLEO are both studying the $\Dsp\to\Km\Kp\pip$ Dalitz plot and have reported preliminary results.  Projections of their fits are shown in Fig.~\ref{fig:kkpidalitz}, and the fit fractions listed in Table~\ref{tbl:kkpidalitz}.  The results agree in general terms, although there is some tension in the region of the $\phi$ due to the scalar contributions (in particular BaBar quotes a large systematic uncertainty in the fit fraction due to $f_0(980)\pip$).  For this purpose the critical question is how much the two descriptions of the Dalitz plot differ on the rate at different points in phase space, which can only be checked when full amplitude and phase information are made available.

\begin{table} 
\caption{\label{tbl:kkpidalitz} Fit fractions of various decay modes in preliminary $\Dsp \to \Km\Kp\pip$ Dalitz analyses from BaBar \cite{Pappagallo:2007it} and \cleoc. All fit fractions are in percent.  The uncertainties on the BaBar results are statistical and systematic, respectively; those on the \cleoc\ results are statistical only.}
\begin{tabular}{ccccccccc}
\hline\hline
Mode & BaBar & \cleoc\\
\hline
$\bar K^{*}(892)^0 \Kp$ & $48.7 \pm 0.2 \pm 1.6$ & $47.4 \pm 1.5$\\
$\bar K_0^{*}(1430)^0 \Kp$ & $2.0 \pm 0.2 \pm 3.3$ & $3.9 \pm 0.5$ \\
$\bar K_2^*(1430)^0 \Kp$ & $0.17 \pm 0.05 \pm 0.3$ & --- \\
$f_0(980)\pip$ & $35 \pm 1 \pm 14$& $28.2\pm 1.9$\\
$\phi\pip$ & $37.9 \pm 1.8$ & $42.2 \pm 1.6$\\
$f_0(1370)\pip$ & $6.3 \pm 0.6 \pm 4.8$ & $4.3\pm 0.6$\\
$f_0(1710)\pip$ & $2.0 \pm 0.1 \pm 1.0$ & $3.4 \pm 0.5$\\
$f_2(1270)\pip$ & $0.18 \pm 0.03 \pm 0.4$ & ---\\
\hline\hline
\end{tabular}
\end{table}

\section{STRONG PHASES FROM QUANTUM CORRELATIONS}
One of the primary techniques for exploring \Dz\ mixing is to look at the time evolution of the ``wrong-sign'' decay $\Dz(t) \to \Kp\pim$, where $\Dz(t)$ indicates a meson, which was produced in the definite flavor eigenstate \Dz, after proper time $t$ has elapsed.  This amplitude receives two contributions:
\begin{itemize}
\item the direct but doubly-Cabibbo-suppressed (DCS) decay $\Dz \to \Kp\pim$,
 \item mixing followed by the Cabibbo-favored (CF) decay $\Dzb \to \Kp\pim$.
\end{itemize}
In the limit of small mixing the first amplitude remains near its initial value, while the second grows linearly; this differing dependence permits a measurement of mixing parameters from the shape of the $\Dz(t) \to \Kp\pim$ rate.  Because of the additional (and larger) amplitude from the DCS decay, the mixing term is amplified by interference and becomes measurable at first order in the mixing parameters.  However the size of the interference depends on the relative phase of the two amplitudes as well as their magnitude, and relating this measurement to the fundamental mixing parameters requires knowing this phase.  The weak contribution to the phase is known, leaving the QCD component to be determined.

This can be done by exploiting the quantum correlations between \D\ pairs at \cleoc, as detailed in Ref.~\cite{Asner:2005wf}.  The process $e^+e^- \to \psi(3770)$ produces a state with the vector quantum numbers $J^{PC} = 1^{--}$.  When this decays to the pseudoscalar pair \Dz\Dzb, the daughters are produced in a $L=1$ angular momentum state, which implies that the product of the daughter $CP$ states must be odd.  We can rotate the basis and think of the $\psi(3770)$ decay as producing a pair of mass eigenstate mesons $D_1 D_2$, which in the absence of $CP$ violation are also $CP$ eigenstates:
\[ |D_{1,2}\rangle = \frac{|\Dz\rangle \pm |\Dzb\rangle}{\sqrt{2}}. \]

One can apply the standard tagging techniques in this new basis.  If a $CP$ eigenstate decay is tagged, this projects out the opposite $CP$ component on the other side --- which is a sum or difference of flavor eigenstates, whose decay amplitudes will interfere.  In particular the resulting total amplitudes (and rates) to common \Dz\ and \Dzb\ final states, for example \Kp\pim, will be dependent on the phase between the two contributing amplitudes.

The \cleoc\ analysis \cite{Rosner:2008fq} to extract the strong phase $\delta$ between $\Dz \to \Kp\pim$ and $\Dzb\to\Kp\pim$ involves reconstructions of three classes of \Dz\ decays: $CP$ eigenstates (e.g.\ $\Km\Kp$ for $CP=+$, $\KS\piz$ for $CP=-$), semileptonic decays (which are unambiguous flavor tags as the lepton has the same charge sign as the parent charm quark), and $K^\mp\pi^\pm$.  The $CP$ correlation modifies the time-integrated double tag yields from what is expected in the ``uncorrelated'' case where the \Dz\ and \Dzb\ decay independently. For example, $CP=+$ versus $CP=+$ is suppressed to zero, while $\Dz\to \Km\pip$ vs.\ $\Dzb \to X e^- \bar\nu_e$ is modified by a factor $(1 - ry \cos \delta -rx \sin \delta)$, where $\delta$ is as above, $r$ is the magnitude of the amplitude ratio between DCS and CF $\Kp\pim$ decays, and $x$ and $y$ are mixing parameters representing the mass and width difference of the mass eigenstates.

In principle enough information is present in this analysis to determine all the input parameters ($r$, $\delta$, and the mixing parameters $x$ and $y$, as well as relevant branching fractions).  In practice \cleoc\ event yields are insufficient to do this, so external inputs are used.  In the standard fit the branching fractions as well as $R_\mathrm{M} \equiv (x^2+y^2)/2$ and the time-integrated ratio $R_\mathrm{WS}\equiv\Gamma(\Dz\to\Kp\pim)/\Gamma(\Dzb\to\Kp\pim)$ are fixed to precise determinations from the $B$-factories and \cleoc.  In addition, in the standard fit $x\sin\delta$ is fixed to zero as it is extremely weakly determined and $\sin\delta$ is expected to be near zero (the associated systematic uncertainty on $\cos\delta$ is $\pm 0.03$).  

\begin{figure}
 \includegraphics[width=0.7\linewidth]{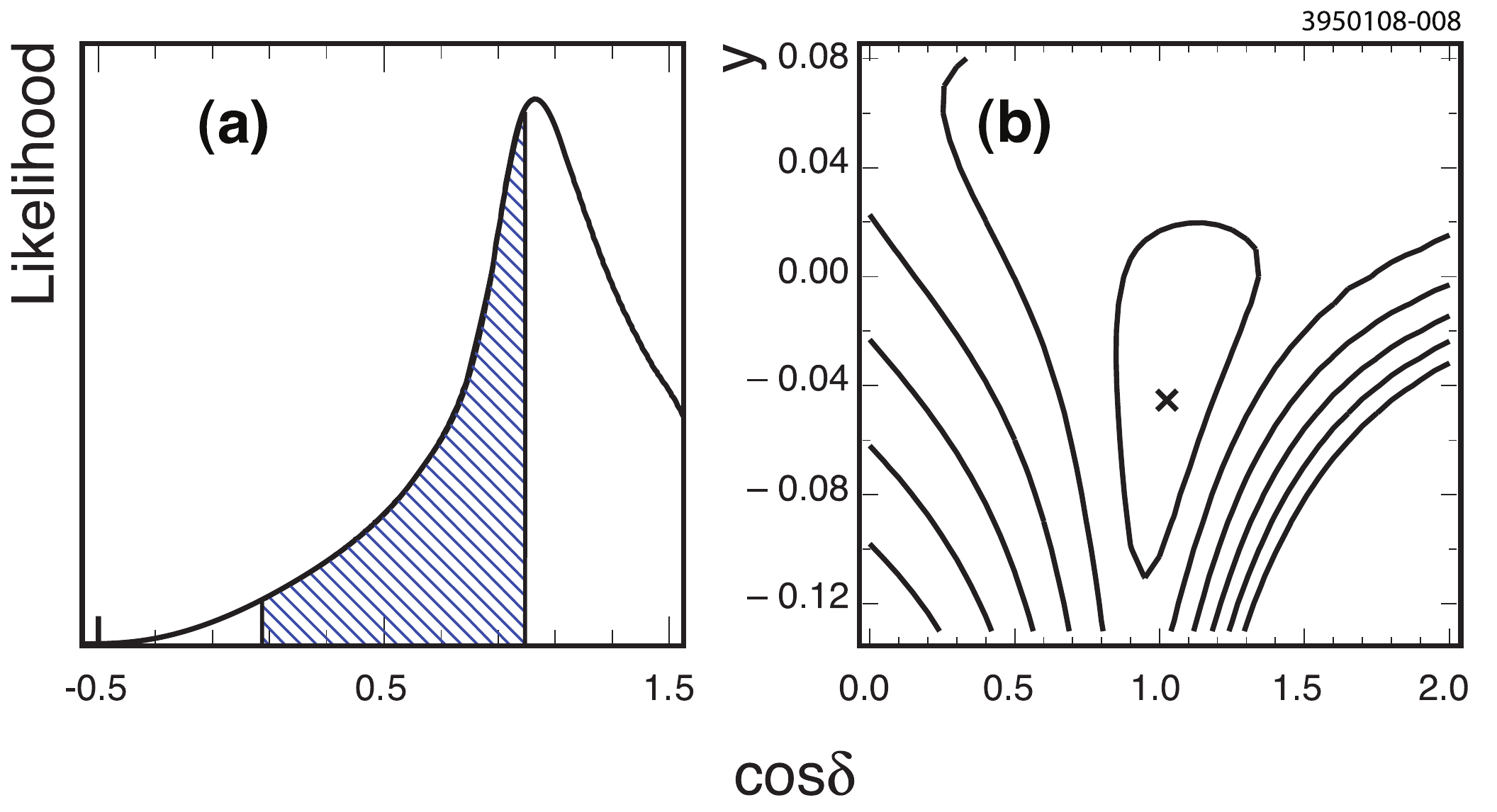}
\caption{\label{fig:stddeltafit}Likelihood function for determination of the $\Km\pip$ strong phase via quantum correlations at \cleoc\ \cite{Rosner:2008fq}.  Plot a) shows the likelihood projected onto the $\cos\delta$ axis; the hatched region shows 95\% of the area in the physical region ($|\cos\delta| < 1$).  Plot b) shows the two dimensional contours in $y$ and $\cos \delta$; the lines are spaced $1\sigma$ apart, where $\sigma \equiv \sqrt{\Delta \chi^2}$. The result $\cos \delta \approx 1$ is stable under excursions in $y$, but the uncertainties are not Gaussian.}
\end{figure}

The likelihood contours from the standard fit are shown in Figure~\ref{fig:stddeltafit}.  The likelihood function is peaked at $\cos\delta \approx 1$, but has a significant fraction of the area outside the physical region.  In addition, the uncertainties are not Gaussian.  By projecting the likelihood onto the $\cos \delta$ axis and integrating 95\% of the area in the physical region, \cleoc\ obtains $\cos \delta > 0.07$ or $|\delta| < 75^\circ$ at 95\% confidence level. This analysis provides the first measurement of $\delta$.

An ``extended'' fit can also be performed incorporating external information on $r$ and the mixing parameters $x$ and $y$.  This significantly reduces the uncertainty on $y$ and permits them to float $x\sin\delta$.  At 95\% confidence level \cleoc\ finds $x\sin \delta \in [0.002,0.014]$ and $\delta \in [-7^\circ,+61^\circ]$.  This result depends on external measurements that can be expected to become more precise, and hence is less useful than the standard fit for incorporation into averages.  

\section{DECAY MECHANISM STUDIES}
Leptonic and semileptonic decays of heavy quarks are much better understood theoretically than hadronic decays, because the relevant weak interactions are demonstrably short distance processes and the daughter leptons do not participate in the strong interaction.  The effects of the strong interactions can thus be absorbed in parameters such as the decay constant for leptonic decays and form factors for semileptonic decays, which can be predicted using various techniques.

In comparison, hadronic decays involve at least four quarks.  These can arrange themselves into hadrons in many ways; in particular the produced hadrons can rescatter into other states.  These ``long distance'' processes occur at the hadronic scale instead of the electroweak scale, and are much less theoretically tractable than short distance interactions and subsequent two-quark behavior.  

Long distance effects can be large.  For example, the short distance contribution to \Dz\ mixing is highly suppressed by the GIM mechanism, but long distance effects due to flavor SU(3) violation can induce significant mixing and are likely the dominant cause of the observed lifetime difference of the two mass eigenstates \cite{Falk:2001hx}.

\subsection{Observation of $\Dsp \to p\bar n$}
The process $\Dsp \to p\bar n$ is the only kinematically allowed decay of a ground state charmed meson to baryons and provides a unique probe of hadron dynamics in weak annihilation decays \cite{Pham:1980xe,Bediaga:1991eu,Chen:2008pf}.  The recently observed large branching fraction of order $10^{-3}$, in contradiction to initial predictions of order $10^{-6}$, raises the question of whether final state rescattering dominates short distance annihilation, or whether applications of the PCAC hypothesis ignoring states heavier than the pion fail in this case.

\begin{figure}
 \includegraphics[width=.4\linewidth]{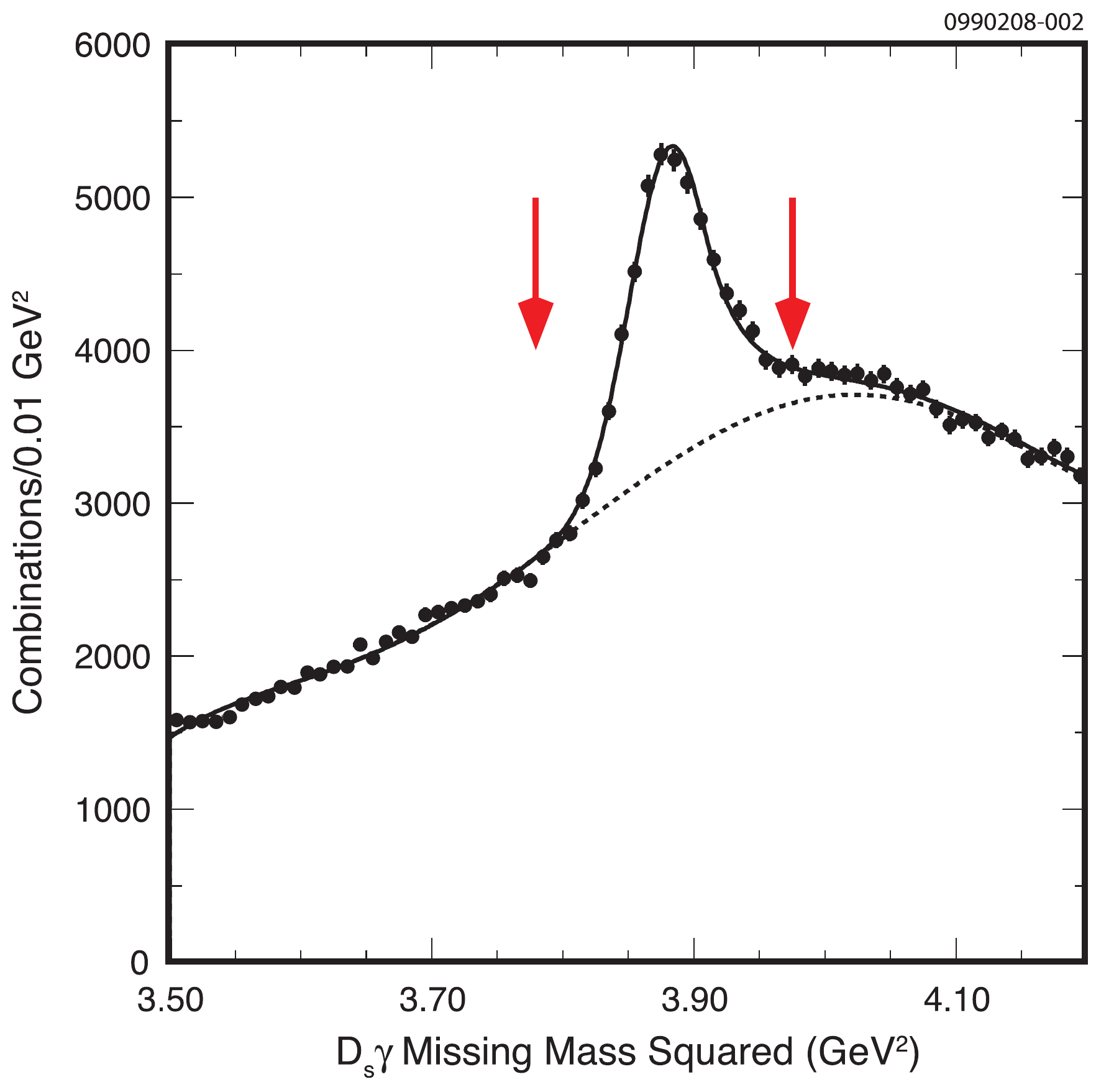}
\includegraphics[width=.4\linewidth]{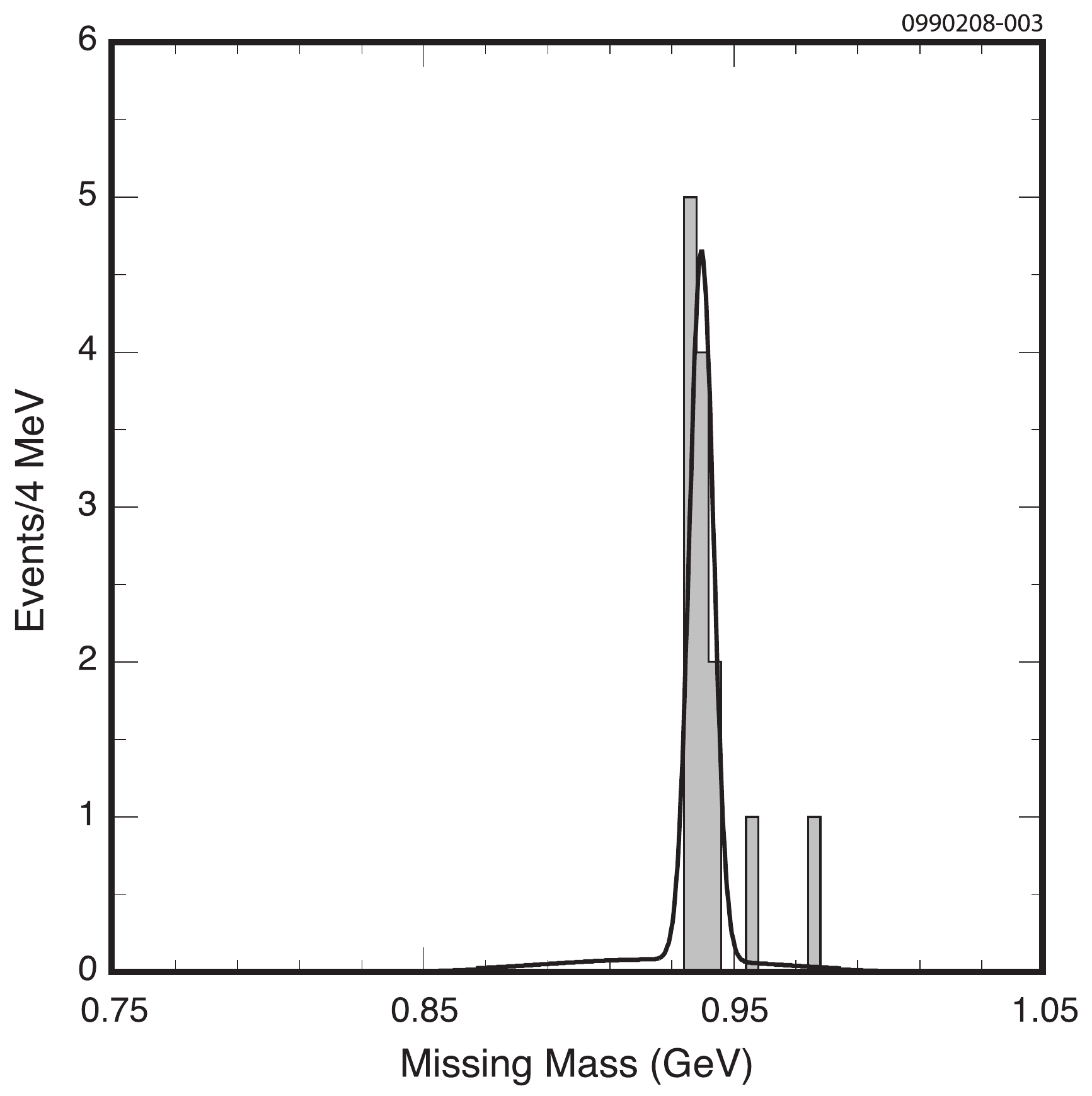}
\caption{\label{fig:dsbaryonic} Observation of $\Dsp \to p\bar n$ at \cleoc\  \cite{Athar:2008ug}.  Left: missing mass squared of tag $\Ds\gamma$ combination, showing peak at $m_{\Ds}^2$ for true $\Ds^{*\pm}\Ds^\mp$ events.  Candidates within the red arrows are then investigated for $\Dsp \to p\bar n$ candidates. Right: missing mass opposite $\Ds \gamma p$ system showing peak at $m_n$.  The solid line shows the expected signal distribution. }
\end{figure}

This decay has been seen for the first time at \cleoc\ \cite{Athar:2008ug}.  No attempt is made to find the neutron or antineutron from the \Dsp\ decay; instead its presence is inferred from kinematic variables, in particular the missing mass opposite the rest of the $\Ds^{*\pm}\Ds^\mp$ event, which is fully reconstructed.  This is analogous to the procedure used for the \cleoc\ measurement of $\Br(\Dsp \to \mu^+ \nu_\mu)$ \cite{Artuso:2007zg} and many of the methods are shared.  A tag $\Dsm$ is reconstructed in a fully hadronic mode, and combined with a photon candidate to see if the missing mass of that combination corresponds to mass of the as far unreconstructed \Dsp.  If so, and in addition the photon is kinematically compatible with arising from the decay of the $\Ds^{*+}$ or $\Ds^{*-}$, a proton track is searched for in the drift chamber; particle identification is done using specific ionization.  If found, the missing mass opposite the $\Dsm\gamma p$ system is evaluated.  Kinematic fits to the assumed intermediate particles are used to improve resolution and reject background.  The results are shown in Fig.~\ref{fig:dsbaryonic}; there are thirteen events in the signal region of 900-980 MeV$/c^2$.  Sidebands in the tag \Dsm\ mass are used to evaluate backgrounds; no sideband events fall in the signal missing mass region.  Thirteen events and no background yields a branching fraction $\Br(\Dsp \to p \bar n) = 1.30 \pm 0.36 ^{+0.12}_{-0.10} \times 10^{-3}$, which is appreciable, especially given the phase space suppression.

\subsection{Two-Body Decay Amplitude Analysis}
Although the inclusion of long distance effects means that short distance Feynman diagrams for weak decays of charmed quarks should not be taken seriously for computational purposes, they are still useful for visualizing the flow of quark flavor during a decay in the ``flavor topology'' approach.  Coupled with flavor SU(3) symmetry, this can be a powerful tool for understanding the relationships between different hadronic decay rates.

\begin{figure}
 \includegraphics[width=.7\linewidth]{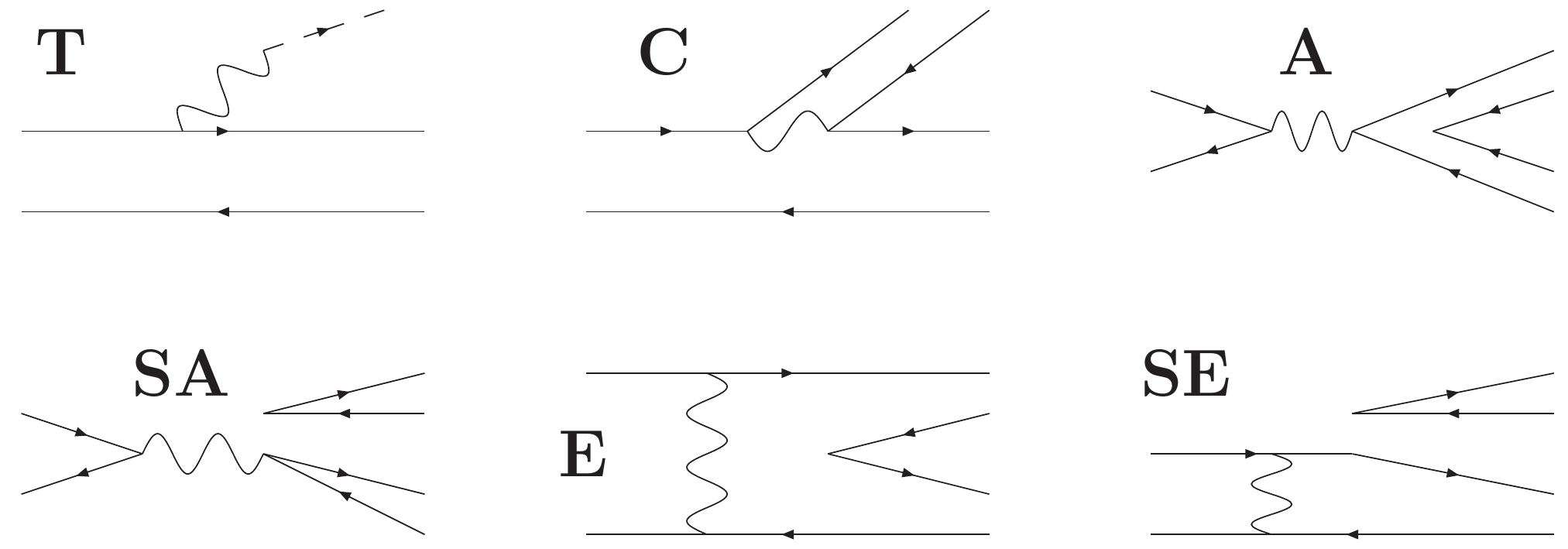}
\caption{\label{fig:rosner_diag} The six flavor topology diagrams used to analyze $D \to PP$ data in Ref.~\cite{Bhattacharya:2008ss}: (top left) tree, (top center) color-suppressed tree, (top right) annihilation, (bottom left) annihilation with singlet emission, (bottom center) exchange, (bottom right) exchange with singlet emission.}
\end{figure}

\begin{figure}
 \includegraphics[height=5.6cm]{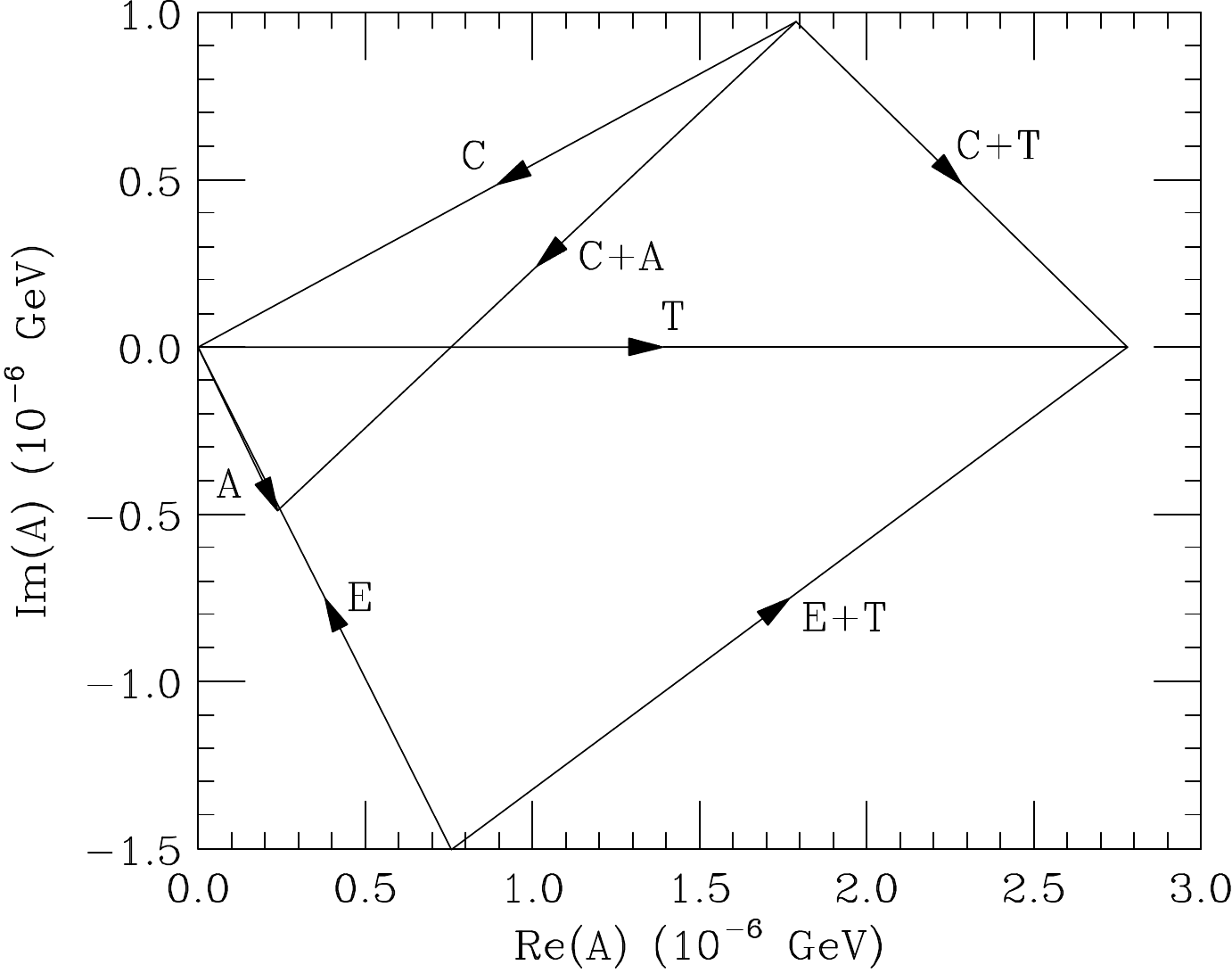}\hskip 0.4cm
 \includegraphics[width=.268\linewidth]{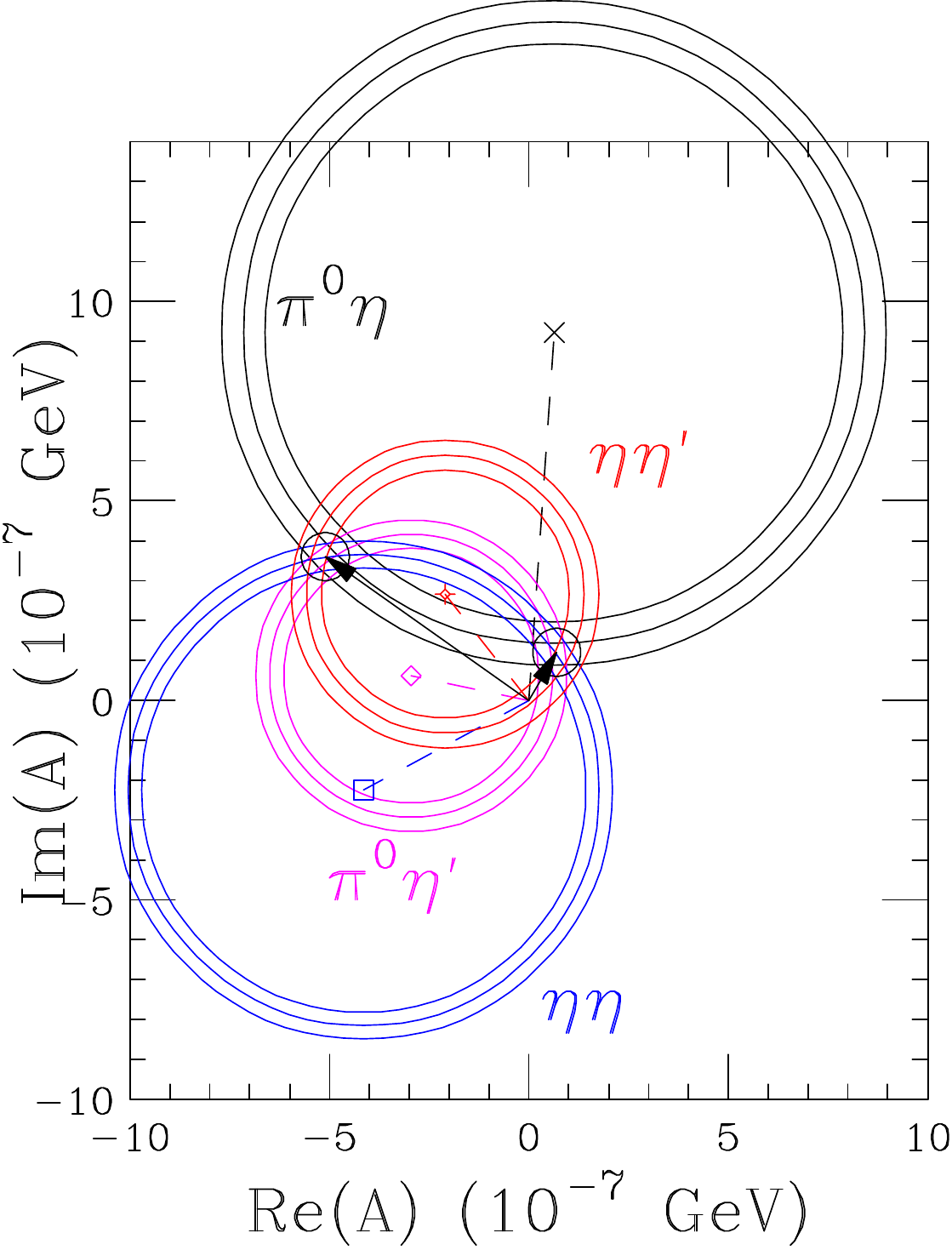}\hskip 0.4cm
 \includegraphics[width=.269\linewidth]{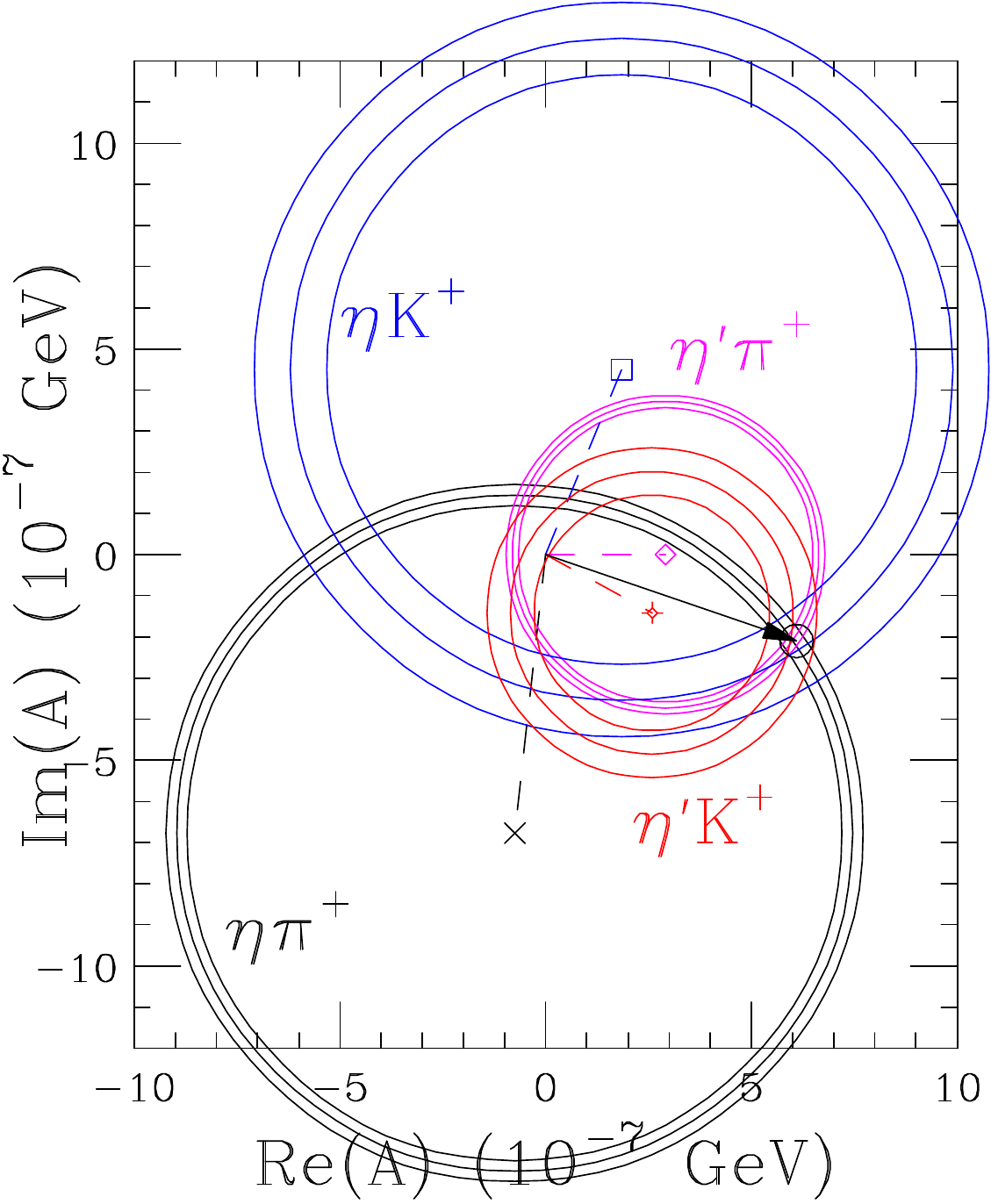}
\caption{\label{fig:pp}Results of analysis of $D\to PP$ decays by Bhattacharya and Rosner \cite{Bhattacharya:2008ss}.  Left: complex amplitude for $T$, $C$, $E$, and $A$ amplitudes (see text and Fig.~\ref{fig:rosner_diag} for explanation of amplitudes).  Center: construction to obtain $SE$ amplitude; allowed solutions lie at the intersection of all circles.  Two solutions are evident.  Right: similar plot for $SA$; there is only one solution.   }
\end{figure}

\cleoc\ has recently assembled a complete picture of Cabibbo-favored and singly-Cabibbo-suppressed open charm decays to pairs of pseudoscalar mesons \cite{Dobbs:2007zt,Alexander:2008cqa,Bonvicini:2008nr,Adams:2007mx,Artuso:2008ri}; many of these are first observations, especially for \Dsp\ decays.  These branching fractions have been analyzed by Bhattacharya and Rosner \cite{Bhattacharya:2008ss} in a comprehensive framework assuming SU(3) symmetry to relate \Dz, \Dp, and \Dsp\ decays.  Six flavor topologies were assumed, and are detailed in Fig.~\ref{fig:rosner_diag}.  The results are shown in Fig.~\ref{fig:pp}.  Among the conclusions are:
\begin{itemize}
 \item the $T$ diagram has the largest magnitude, but $C$ and $E$ are also fairly large: $|C| \sim 0.75 |T|$, $|E| \sim 0.6 |T|$;
\item there are large phases between $T$, $C$, and $E$ (e.g. 150$^\circ$ between $C$ and $T$);
\item the annihilation diagram is significantly smaller and approximately antiparallel to $E$: $A \sim -0.3E$;
\item there are two allowed solutions for $SE$, both of which are small compared to $E$;
\item there is one solution for $SA$, which is of comparable magnitude to $A$.
\end{itemize}
It should be emphasized that these conclusions are only valid for two-pseudoscalar decays, and that because of long distance effects these should not be interpreted as statements about short distance physics.  (For example, the large $E$ amplitude does not imply that short distance $W$ exchange amplitude is also large.)

\section{DALITZ PLOT STUDIES}
The decays of $D$ mesons provide interesting information on light hadron properties.  All possible routes to the same final state interfere, so phase information is available.  Because the initial state is a pseudoscalar, the angular distributions are highly constrained.  The case of decays to three pseudoscalars is particularly popular to study (``Dalitz analysis''), as there are only two independent kinematic variables and no information hidden in the final state daughter spins.  In general it is assumed that the variation of the amplitude for the decay $X \to ABC$ can be decomposed into a the coherent sum of amplitudes for a few intermediate states, where two of the daughters arise from a resonance and the third is a spectator (for example, $\Dsp \to \pip\pip\pim$ could have contributions from $\Dsp \to f_0(980)\pip$, $f_2(1270)\pip$, and $\rho(770)\pip$); this is termed the isobar model.  In particular all amplitude and phase variation is assumed to come from the intermediate resonance which is generally parametrized using Breit-Wigner or Flatt\'e functions.

This description, while very useful, has recently been recognized as inadequate, especially for describing the $S$-wave component of decay amplitudes.  In particular a coherent sum of interfering Breit-Wigner or Flatt\'e lineshapes does not conserve unitarity, and the resonances in the scalar sector are so wide that this is a major issue.  There is also a debate on the nature of the scalar sector below 1 GeV$/c^2$, where structures referred to as the $\sigma$ and $\kappa$ are suggested to contribute to $\pi\pi$ and $K\pi$ scattering, respectively.  Their importance in charm decays is unclear.  In the absence of more sophisticated techniques, the $S$-waves in $D$ decays are typically parametrized as constant (``non-resonant'') amplitudes with ad hoc ``resonances'' added where the non-resonant description fails drastically.

One technique for unitarizing the $S$-wave amplitudes, called the $K$-matrix approach, has been successfully applied to charm decays in the past.  It relies on outside information on phases and amplitudes of low-energy scattering to derive amplitudes, and is explicitly unitary.  Unfortunately this method is complex, the physical meaning of the results is usually not especially transparent, and it requires external input --- the experimental result is not self-contained.

Recently another approach, reliant on high statistics, has been pioneered by the E791 Collaboration \cite{Aitala:2005yh}: the ``model-independent partial wave analysis'' (MIPWA).  If two amplitudes with different angular dependence interfere, and a model (Breit-Wigner, etc.) is available for the phase and magnitude of one, the complex amplitude for the other can determined, up to a twofold phase ambiguity, without assumptions on its energy dependence.  (Requiring smooth evolution of the amplitude with energy can help choose between solutions.)  Vector and tensor resonances in $D$ decays are generally modeled well by Breit-Wigner lineshapes with well-known masses and widths, which can be fixed in order to determine the amplitude and phase of the $S$-wave components.  Ambiguities can still arise in such an approach but the fact that the raw amplitudes are made available makes this a very powerful technique.

Here I mention two such analyses.  The first is a \cleoc\ study of the $D^+ \to \Km\pip\pip$ decay \cite{Bonvicini:2008jw} using $141 \times 10^3$ events.  The dominant contributor is found to be the $\Km\pip$ $S$-wave.  The standard MIPWA is modified by explicitly including a $\overline{K}_0^{*}(1430)^0 \pip$ contribution with a Breit-Wigner shape for the $\overline{K}_0^{*}(1430)^0$; this is done so that the remaining scalar amplitude varies slowly. This amplitude is observed to be large at low $\Km\pip$ mass; however it does not follow the expectation for a $\kappa$ alone, and can only be modeled in that framework by adding a large non-resonant component that interferes significantly with the $\kappa$ at low mass.  From this it is unclear that the $\kappa$ is a compelling model for the $\Km\pip$ $S$-wave in this decay.

\begin{figure}
 \includegraphics[height=7cm]{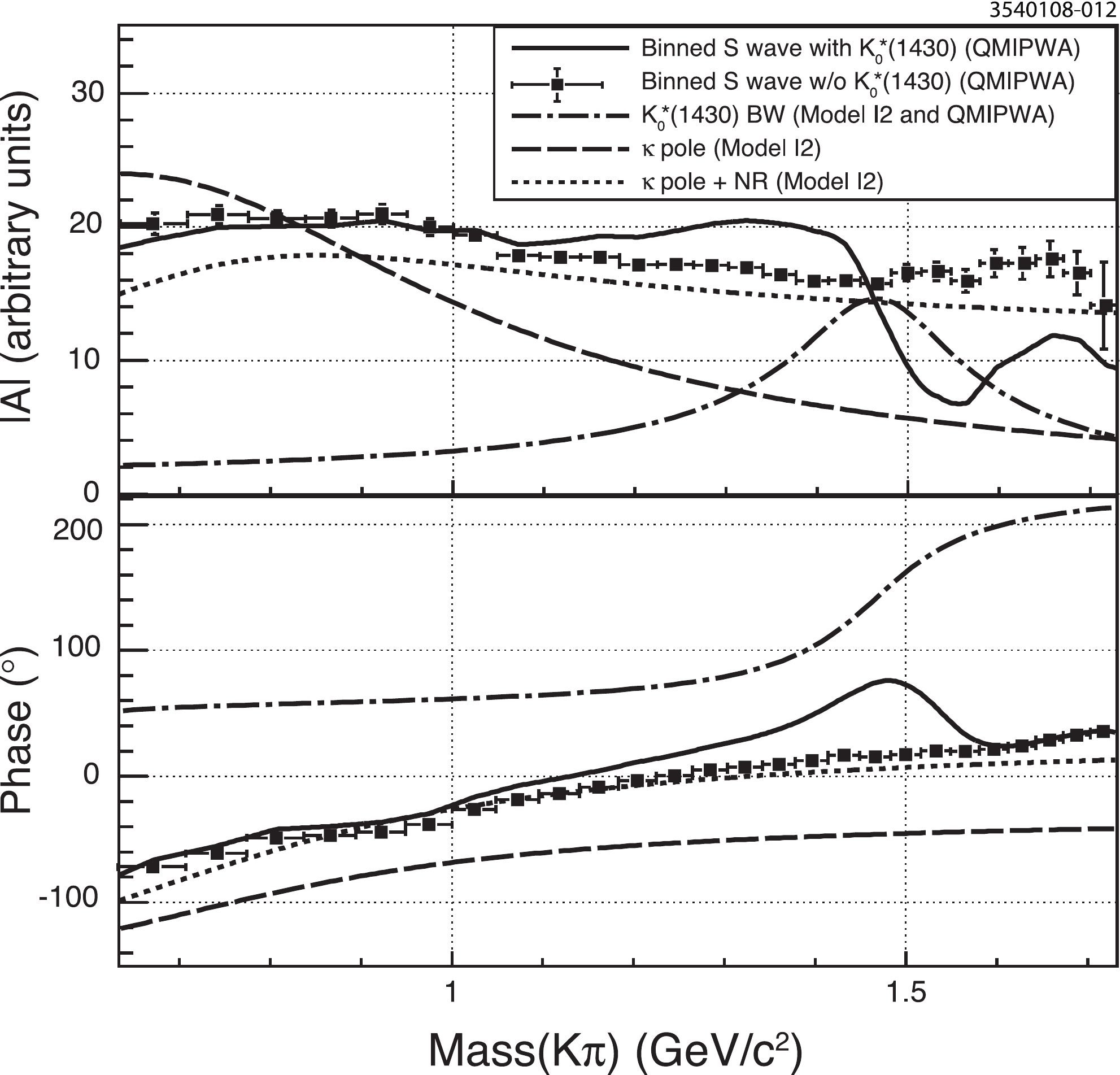}
 \includegraphics[height=7cm]{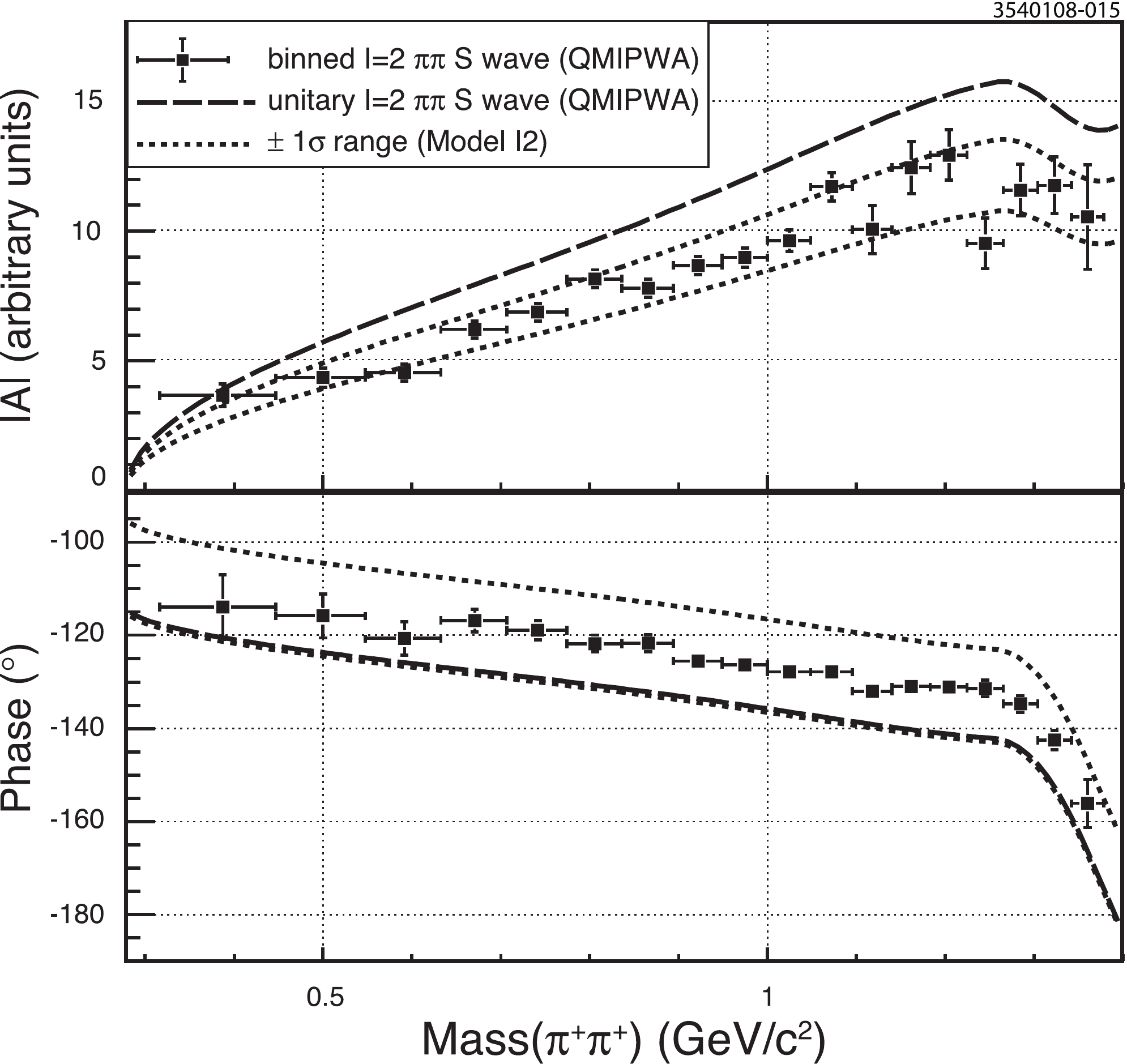}
\caption{\label{fig:kpipipwa}Scalar $\Km\pip$ (left) and $\pip\pip$ (right) amplitudes in the process $\Dp \to \Km\pip\pip$, as obtained by \cleoc\  \cite{Bonvicini:2008jw}.  For the $\Km\pip$ plot, the points indicate the MIPWA results for the amplitude excluding the $K_0^*(1430)$ contribution; the dot-dashed line shows the $K_0^*(1430)$ Breit-Wigner shape; the solid line indicates the sum of the MIPWA and $K_0^*(1430)$ amplitudes; the dotted line shows the result of an isobar model fit including a $\kappa$ and a constant nonresonant $\Km\pip$ amplitude; the dashed line shows the $\kappa$ component of that fit.  For the $\pip\pip$ plot, the points show the result of the MIPWA fit for the $\pip\pip$ amplitude; the dashed line shows the analytic amplitude used in the MIPWA fit for the $\Km\pip$ amplitude; the dotted lines show $\pm 1\sigma$ variations on the analytic amplitude used in the isobar model fit to the $\Km\pip$ amplitude. }
\end{figure}

The fit also requires an appreciable $\pip\pip$ $S$-wave component (fit fraction 10--15\%), whose parametrization is taken from fits to pion scattering data and which has significant variation with $\pip\pip$ invariant mass.  The MIPWA technique was used to obtain amplitudes for this component as a cross-check of the nominal analytic function, and the two show good qualitative agreement.  The results for the $\Km\pip$ and $\pip\pip$ $S$-wave amplitudes are shown in Fig.~\ref{fig:kpipipwa}.

The second is a preliminary BaBar analysis of the $\Dsp \to \pip\pip\pim$ decay using $13\times 10^3$ events, discussed in more detail in Ref.~\cite{Aubert:2008tm}.  This decay is dominated (83\% fit fraction) by the $\pip\pim$ $S$-wave.  For the \Dsp\ these studies are particularly compelling because of the large fraction of hadronic decays where the Cabibbo-favored $s\bar s$ pair is not manifest, where one mechanism to explain this would involve long distance $s\bar s \to u\bar u$, $d\bar d$ through states where those mix, such as the scalar sector.  The only large non-scalar contribution is from $f_2(1270)\pip$.

\begin{figure}
 \includegraphics[height=7cm]{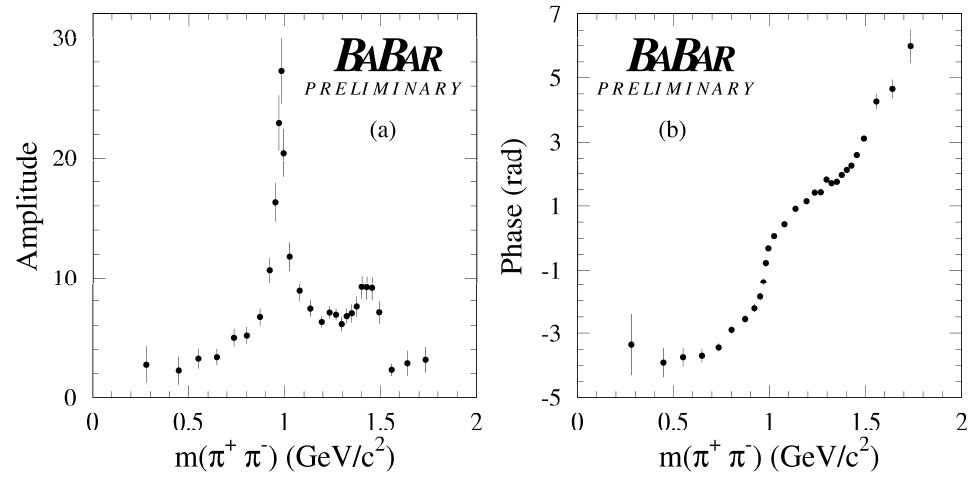}
\caption{\label{fig:3pipwa}Scalar $\pip\pim$ amplitude in the decay $\Dsp\to\pip\pip\pim$ from a preliminary BaBar analysis \cite{Aubert:2008tm}, showing large $f_0(980)$ contribution and structure in the 1.4 GeV$/c^2$ region.}
\end{figure}

The results of the MIPWA are shown in Fig.~\ref{fig:3pipwa}.  There is a clear peak for the $f_0(980)$ with the expected variation in phase for such a dominant resonance.  In the 1400 MeV$/c^2$ region, there is also significant activity.  However, the low mass amplitude or phase variations are not drastic, and careful study is needed to gauge the compatibility of these results with significant $\sigma$ production.

\section{SUMMARY AND OUTLOOK}
In the last few years, the large datasets that have become available at \cleoc, BaBar, and Belle have enabled sophisticated studies of hadronic \D\ decays to be performed.  These results have reduced branching fraction uncertainties by factors of 2 to 4, provided first measurements of strong interaction phases between two body decays, improved our understanding of decay dynamics, and shed light on light hadrons.  There is more to learn in all these areas.  The high-luminosity BES-III experiment has recently started taking data and will have much to contribute to our understanding of the hadronic decays of open charm.

\begin{acknowledgments}
I thank D.~Asner, M.~Dubrovin, S.~Prell, J.~Rosner, Y.~Sakai, M.~Shepherd, and W.~Sun for help and invaluable discussions.  This work is partly supported by a Fermi Fellowship from the University of Chicago.
\end{acknowledgments}


\end{document}